\documentclass[a4paper,aps,prx,superscriptaddress, nofootinbib]{revtex4-2}

\usepackage{xcolor} 

\usepackage{hyperref}
\hypersetup{
    colorlinks,
    linkcolor={blue!90!black},
    citecolor={blue!10!blue},
    urlcolor={blue!80!black}
}
\usepackage{dirtytalk}
\usepackage{tikz}
\usepackage{amssymb}
\usepackage{amsmath}
\usepackage{graphicx}
\usepackage{dblfloatfix}
\usepackage{multirow}

\usepackage{soul}
\usepackage{lineno}

\usepackage[normalem]{ulem}

\begin{document}

\title{DNA Replication under Thermal, Chemical, and Genotoxic Stress}

\author{Chinmaya Pradhan}
\email{chinmaypradhan5000@gmail.com}
\affiliation{Department of Physics, Indian Institute of Technology Hyderabad, Hyderabad 502284, India}

\author{Bhakti Mehta}
\affiliation{Tata Institute of Fundamental Research, Hyderabad 500046, India}

\author{Nirjharini Saha}
\affiliation{Tata Institute of Fundamental Research, Hyderabad 500046, India}

\author{Mrinal Srivastava}
\email{mrinals@tifrh.res.in}
\affiliation{Tata Institute of Fundamental Research, Hyderabad 500046, India}

\author{Anupam Gupta}
\email{agupta@phy.iith.ac.in}
\affiliation{Department of Physics, Indian Institute of Technology Hyderabad, Hyderabad 502284, India}

\begin{abstract}
Eukaryotic DNA replication must remain robust under thermal, chemical, and 
genotoxic stress despite large fluctuations in replication dynamics. Here, we 
develop a lattice-based stochastic Monte Carlo framework for whole-genome 
replication in \textit{Saccharomyces cerevisiae} at single base-pair resolution, 
incorporating probabilistic origin firing, replication fork-speed distributions, 
and a time-dependent limiting factor that governs the availability of cellular 
replication resources. The model is benchmarked quantitatively 
against experimental replication profiles before being applied to stress 
conditions, and reproduces diverse replication stress responses using only two 
effective parameters. Importantly, the analysis reveals that replication 
fork-speed heterogeneity underlies the emergence of Erlang-distributed S-phase 
durations and rare, anomalously prolonged replication events observed 
experimentally in \textit{Escherichia coli} and human cell lines, while 
predicting similar behavior in \textit{S. cerevisiae}. The framework further 
predicts non-monotonic thermal behavior, power-law scaling under hydroxyurea 
stress, and total replication-time dynamics under diverse genotoxic conditions.
\end{abstract}

\maketitle

\section{Introduction}
Faithful genome duplication is a fundamental process required to ensure the accurate transmission of genetic material. In eukaryotic cells, DNA replication is spatio-temporally regulated, where synthesis occurs during the S-phase of the cell cycle \cite{zhou2021stochasticity}. DNA replication must adapt to a dynamic intracellular environment and proceed with high fidelity to preserve genomic integrity. Compromised replication can result in genomic instability due to the stall/collapse of the replication fork, changes in the ploidy, or cell death \cite{willis2009regulation, gaillard2015replication}.

In eukaryotic cells, DNA replication initiates at locations known as origins of replication, whose selection mechanisms vary across species. Replication origin selection in eukaryotes varies across species, being largely governed by chromatin and epigenetic context in mammals and \textit{Xenopus}, associated with AT-rich regions in fission yeast, and directed by sequence-defined autonomously replicating sequences (ARSs) in budding yeast. Owing to their large size, eukaryotic genomes rely on simultaneous activation of multiple replication origins to complete DNA replication in a timely manner \cite{sekedat2010gins,berners2025regulation}. The DNA double helix undergoes melting at the origin, giving rise to bidirectional Y-shaped structures referred to as replication forks.  The global progression of the S phase is governed by two key processes: the spatiotemporal activation of replication origins and the speed of fork progression. Tightly coordinated control of these two processes is suggested to be mediated by the molecular interplay of regulatory cell-cycle and replication stress response proteins.

 Budding yeast (\textit{Saccharomyces cerevisiae}; \textit{S. cerevisiae}) represents one of the simplest and most tractable systems for studying eukaryotic DNA replication, owing to its short cell cycle and relatively simple, well-characterised genome \cite{berners2025regulation, goffeau1996life}. Replication in \textit{S. cerevisiae} initiates from specific DNA sequences on the genome known as Autonomously Replicating Sequences (ARSs). An ARS consists of an AT-rich core A-element termed ARS Consensus Sequence (ACS) and other auxiliary elements (B1, B2, B3), which together aid in binding of the origin licensing proteins to form the pre-replicative complex (pre-RC) during the G1 phase of the cell cycle \cite{breier2004prediction}. During licensing, the inactive Mcm2-7 double hexamer helicase complex is loaded onto the ARSs throughout the genome, marking potential origins, a subset of which is activated/fires during the S phase and initiates DNA replication \cite{chang2011high}. Recent evidence supports the presence of a much higher pool of potential origins at the end of G1, as compared to the subset that ultimately fires and contributes to the replication of the yeast genome during the S phase \cite{foss2024identification}. The majority of the origins that do not fire (dormant origins) are thought to serve as a backup for the rescue of DNA replication under genotoxic stresses such as DNA damage, nucleotide depletion, or environmental stress\cite{dukaj2021capacity}.

Upon entry into S phase, phosphorylation of Mcm2–7 and firing factors Sld2/Sld3 by DDK and S-CDK promotes recruitment of Cdc45 and GINS to form the active CMG helicase, which travels with replication forks \cite{labib2010cdc7,de2021ddk,kohler2016cdc45,yuan2020dna}.
In contrast, DDK and S-CDK act transiently at origins and diffuse after origin activation\cite{rhind2022dna,patel2008hsk1,mantiero2011limiting}, while CMG is released only upon fork convergence \cite{xia2021fate}.
These observations suggest that origin activation and fork progression depend on a finite pool of replication-associated resources that are recycled after fork merger.  Motivated by this regulatory mechanism, we introduce limiting factors that act as effective cellular resources to trigger origin firing, thereby generating two active replication forks that drive fork progression. Upon the merger of two converging forks, the resources responsible for fork movement are released and subsequently recruited by other inactive origins to promote their firing.
These cellular resources, therefore, limit the number of replication forks that can be active at any given time, providing a global regulatory mechanism for origin firing. Motivated by earlier work in human DNA replication, in which the availability of limiting factors was described using a saturating time-dependent function by Lob et al.\cite{lob20163d}, we introduce a \textit{S. cerevisiae }- specific limiting factor whose temporal dynamics are constrained by experimentally measured active fork statistics.

Other than the regulatory function of the limiting factor, fork speed distribution also plays a major role in the replication program. Replication fork progression in budding yeast is not uniform but spans a broad distribution of speeds across the genome, reflecting local chromatin context, transcriptional activity, and nucleotide availability \cite{theulot2022genome}. Slower forks often arise at protein-DNA barriers, actively transcribing regions, or compact chromatin, allowing time for repair and chromatin maturation, whereas faster forks occur in more permissive regions \cite{azvolinsky2009highly}. This variability helps coordinate origin firing and prevents excessive fork density that could lead to collisions or depletion of replication machinery \cite{mantiero2011limiting}. Under stress conditions, regulation of fork speed contributes to checkpoint activation and genome stability maintenance. Thus, fork speed heterogeneity is thought to be a key regulatory layer ensuring robust completion of S phase while minimizing replication-associated DNA damage \cite{theulot2022genome}.

\textit{S.cerevisiae} was conventionally considered to exhibit a predominantly deterministic DNA replication program \cite{berners2025regulation, pellet2025replication, labit2008dna, mantiero2011limiting} since its initiation sites and their corresponding firing times have been well characterized by high-throughput experiments\cite{hawkins2013high, gilbert2010evaluating}. A deterministic replication program would imply that pre-defined, site-specific origins fire in almost every cell cycle as per its firing efficiency, in contrast to fission yeast or higher eukaryotes, in which DNA replication initiates asynchronously throughout the S-phase at random sites \cite{barberis2013replication}. However, extensive experimental evidence demonstrates substantial variability in origin firing efficiency \cite{friedman1997replication, tuduri2010defining} that can be well captured by a stochastic firing model \cite{hawkins2013high}.

In addition to the variability in replication origin activation, the distribution of inter-origin distances exhibits an approximately exponential decay \cite{hawkins2013high}. Such a distribution is consistent with either clustered origin firing or an inherently stochastic origin firing process \cite{kelly2019dynamics,patel2006dna}. These observations motivate us to formulate a stochastic model of origin firing in S. cerevisiae. Additionally, existing models are typically highly constrained processes with multiple parameters while presuming a constant mean speed. The assumption of constant fork speed can substantially influence how the total replication times are distributed across thousands of simulations, each representing an independent experiment. Therefore, a quantitative framework that integrates stochastic origin firing with fork speed distribution is essential not only to accurately reconstruct organismal replication but also to understand the principles of its spatiotemporal adaptations under various cellular and environmental stresses. Diverse thermal, chemical, and genotoxic stress conditions in yeast have shown to modulate both origin firing efficiency and replication fork speed, highlighting that genome replication is a highly plastic process shaped by probabilistic origin activation rather than deterministic origin usage \cite{vanoni1984effects,poli2012dntp,theulot2022genome}.

Here, we introduce a lattice-based stochastic framework for eukaryotic DNA replication at the finest possible spatial resolution (1 bp), in which genome-wide replication dynamics are governed by two experimentally motivated parameters: a time-dependent limiting factor required for origin firing and a fork speed distribution. Applying this model to S. cerevisiae, we demonstrate how rare, prolonged total replication times of cells are dictated by the fork speed distribution. We further investigate how thermal, chemical, and genotoxic stress modulate these two parameters and thereby reshape replication dynamics. Our results suggest that robust genome replication across diverse conditions can emerge from this minimal framework, providing a unifying principle for replication control.

\section*{\textbf{Theoretical Model}}
\subsection*{Architecture of the Base-Pair–Resolution Replication Model}
\begin{figure}[h]
\centering
\includegraphics[width=1\linewidth]{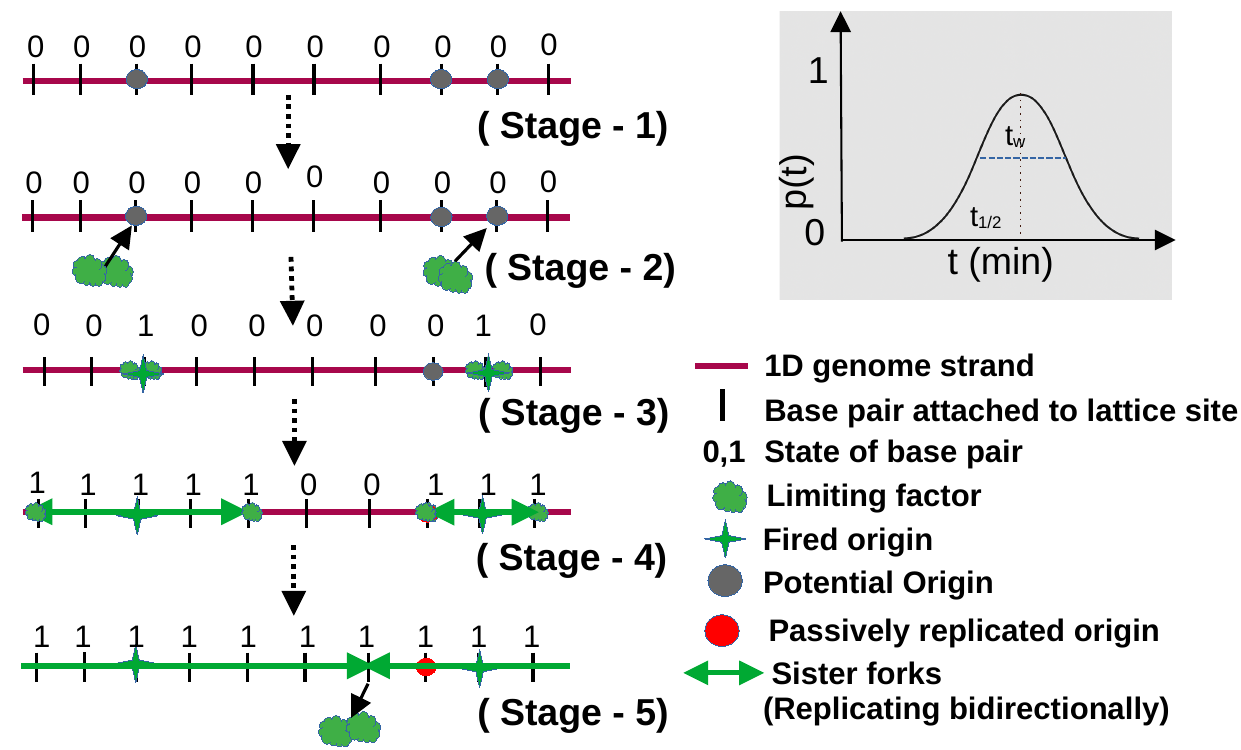}
\caption{Schematic of the model for budding yeast. (Stage 1) Base pairs are assigned to lattice sites along a 1D genome strand and are initially marked as an unreplicated state (‘0’). Potential origins are positioned at predefined genomic locations. The temporal firing probability of each potential origin is represented by a Gaussian distribution with mean ($t_{1/2}$) and variance ($t_{w}$), shown in the top-right shaded panel. (Stage 2) Depending on the availability of the limiting factor $L(t)$, a pair of factors is recruited to a randomly chosen potential origin. (Stage 3) Under suitable conditions, as discussed in the main text, the selected potential origin fires. (Stage 4) A fired origin produces a pair of sister forks that propagate bidirectionally. Replication proceeds behind the forks, converting the base-pair states from ‘0’ to ‘1’. A potential origin undergoes passive replication if it has not fired before an incoming fork replicates it. Each fork is guided by a single limiting factor. (Stage 5) Forks continue to replicate base pairs ($0\rightarrow 1$) before converging with an opposing fork, a point at which the pair of limiting factors is released.}
\label{fig:schematic}
\end{figure}

In our model, the genome is represented as an unpacked one-dimensional strand formed by linear, successively connected chromosomes. Each base pair is mapped onto a lattice site along the strand, and is assigned a binary state: 0 (unreplicated) or 1 (replicated) (Fig.~\ref{fig:schematic}). Multiple replication origins fire both simultaneously and temporally across the genome during S phase, according to their firing probability distribution. 

 For simplification, chromosome boundary conditions are omitted at the whole-genome level. Hence, an active replication fork does not terminate at chromosome boundaries, but continues until it merges with an opposing fork. The region replicated by forks from neighbouring chromosomes rather than by forks from the same chromosome is negligible relative to the length of an individual chromosome. Therefore, this assumption does not influence our results. Hence, this linearly connected chromosome representation is sufficient to investigate the statistics of spatiotemporal origin firing and replication-fork dynamics, while deliberately neglecting three-dimensional structural and biological complexities.

\subsection*{Temporal Probability Distribution of Origin Firing} Our model utilizes experimentally determined information of 459 potential origins. The efficiency of a potential origin to fire is characterized by three properties: (i) competence (P), the normalized proportion of cells in which an origin fires; (ii) median firing time ($t_{1/2}$); and (iii) width ($t_{w}$) of the firing probability distribution curve with time. The quantitative values of these parameters are taken directly from the supplementary table provided by Hawkins et al. \cite{hawkins2013high} and used as inputs in our modeling framework . At the initial stage (stage 1; Fig.~\ref{fig:schematic}) of S phase, each base pair is initialized to `0' (unreplicated), and potential origins are positioned according to their genomic coordinates, which are also taken from the supplementary table of Hawkins et al. \cite{hawkins2013high}. The firing-time distribution is modeled as a Gaussian function, $G(t)$. A schematic of the time-dependent firing probability distribution p(t) of a potential origin is shown in the top-right shaded panel of Fig.~\ref{fig:schematic}. The probability of a potential origin firing at time t is 
\begin{equation}
    p(t) = P \times G(t),
\end{equation}

where
\[
    G(t) = \exp\!\left( -\frac{(t - \mu)^2}{2\sigma^2} \right), \qquad
    \mu = t_{1/2}, \qquad
    \sigma = \frac{t_w}{2}.
\]
\subsection*{Formulation of Limiting factors}Each active replication fork requires one limiting factor; hence, the number of available limiting factors L(t) follows the experimentally observed number of active forks F(t), until it reaches a maximum value during the S-phase. To parameterize the temporal dynamics of active replication forks up to their maximum value, we extracted characteristic features from previously published experimental measurements of time-dependent active forks number by Hawkins et al. \cite{hawkins2013high}. Specifically, we quantify (i) the maximum (peak) value of the time-dependent active fork profile, (ii) the initial growth rate, and (iii) the characteristic time at which the fork number reaches half of its maximum value. Guided by these experimentally measured quantities, we model the time-dependent availability of the limiting factors $L(t)$ as a sigmoid function given by
\begin{equation}
    L(t) = L_{\max} \left( \frac{1}{1 + \exp\!\left[-\beta (t - \tau)\right]} \right)
\end{equation}
where $L_{max}$ denotes the maximum number of available limiting factors, $\beta= 0.4$ controls the growth rate (steepness) of L(t) and $\tau=15$ represents the characteristic midpoint time at which $L(t)=\frac{L_{max}}{2}$. The values of $\beta$ and $\tau$ are kept identical across all stress conditions, whereas  $L_{max}$ is varied. An origin can initiate replication only when two limiting factors are available to create two active forks from that origin (stage-2: Fig.~\ref{fig:schematic}).

\subsection*{Regulatory Conditions for Origin Activation}A potential origin fires only when following conditions are satisfied:
(i) The randomly sampled origin must not have fired earlier.
(ii) The origin must not have been passively replicated by any fork;
(iii) Two limiting factors must be available at the time of firing, and
(iv) The time-dependent firing probability must satisfy a Monte Carlo sampling criterion: at a given time, an origin fires only if its firing probability exceeds a random number drawn from a uniform distribution between 0 and 1. Once an origin meets those suitable conditions and fires, two limiting factors remain occupied onto the active forks generated from that origin, and the state of the corresponding base pair changes from the unreplicated state `0' to the replicated state `1' (stage-3: Fig.~\ref{fig:schematic}). 
Because the combined length of the 16 chromosomes is large, purely stochastic origin sampling can generate long inter-origin gaps, leading to delayed completion of genome replication even when individual origins satisfy the firing criteria. To suppress the formation of such large gaps, we partition the genome into two equal halves, each comprising eight chromosomes. An origin is first sampled stochastically from chromosomes 1–8 and tested for firing, if it does not satisfy the firing conditions, a second origin is sampled from chromosomes 9–16~\cite{rhind2006dna}. The stochastic origin sampling rate `R' is assumed to be constant throughout the S phase and is set as a prefactor (0.06) multiplying the total number of potential origins. The prefactor is inferred by statistically matching the model predictions with experimental data, as detailed in the Supplementary Information.

\subsection*{Formulation of Fork speed distribution}Once an origin fires, it creates a pair of sister forks that move bidirectionally, converting the state of base pairs from `0' to `1' (stage-4: Fig.~\ref{fig:schematic}). Although sister forks from the same origin move at the same speed `$v$' as discussed by Sekedat et al.\cite{sekedat2010gins}, the sister forks generated from other origins may have different speeds, depending on the fork speed distribution `$f(v)$'. Analysis of the experimental fork-speed distribution shape \cite{theulot2022genome} indicates that the distributions can be consistently described by a skewed Gaussian form, given by
\begin{equation}
f(v) = \frac{1}{\omega \sqrt{2\pi}}
\exp\!\left[-\frac{(v-\xi)^2}{2\omega^2}\right]
\left(1 + \operatorname{erf}\!\left(\frac{\alpha (v-\xi)}{\omega \sqrt{2}}\right)\right)
\end{equation}
The skewed Gaussian distribution is parameterized by a location parameter ($\xi$),  width ($\omega$), and skewness ($\alpha$). However, the experimental data only provide the mean fork speed ($\mu$) and width ($\omega$). Therefore, we fix $\mu$ and $\omega$ and assume an approximate value for the skewness parameter ($\alpha$) at $25^\circ\mathrm{C}$ and $30^\circ\mathrm{C}$ (Fig.~\ref{fig:S2}). The location parameter ($\xi$) is then determined such that the resulting skewed Gaussian distribution reproduces the experimentally observed mean. Random variates from the skew-normal distribution were generated using the standard Azzalini construction, based on linear combinations of independent Gaussian variables. Those random variates are assigned to the sister forks to determine their fork speeds, which were held constant until the fork merge. Incorporating this experimentally measured fork-speed distribution into our model enhances its biological realism and strengthens its consistency with experimental observations.

\subsection*{Replication Dynamics Using the Model}As multiple active forks replicate the genome simultaneously, a fork will eventually encounter another fork approaching from the opposite direction. Such an event is called merging of forks, or a termination event, which leads to the release of the two occupied limiting factors (stage-5: Fig.~\ref{fig:schematic}). Now these two limiting factors, recycle and available to fire another origin. Subsequently, DNA replication proceeds through origin firing, fork formation, and the replication of base pairs by active forks. The simulation is terminated once the total replicated fraction reaches 99\% of the genome length, consistent with experimental observations by Hawkins et al. \cite{hawkins2013high}, which show saturation of the replicated fraction curve near this point; the corresponding time is defined as the total replication time ($T_{rep}$) or the total S-phase duration.
\begin{figure*}[t]
\centering
\includegraphics[scale=0.7]{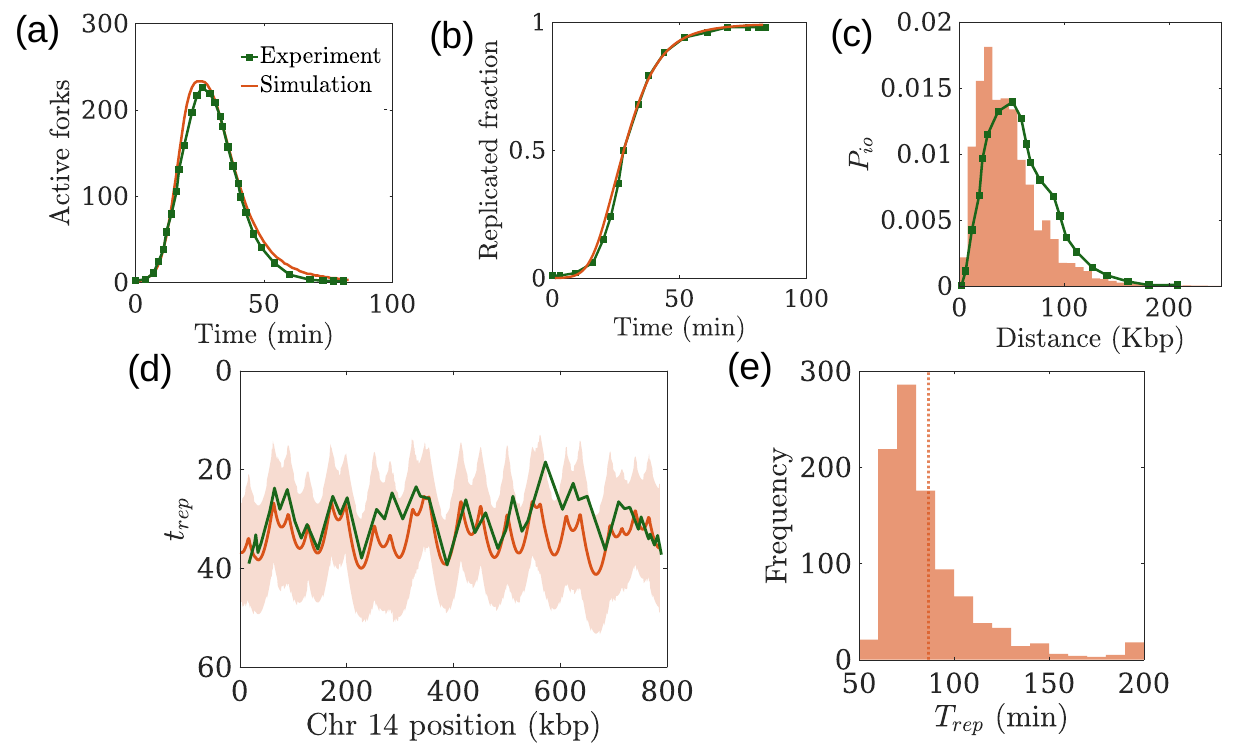}
\caption{Comparison of replication dynamics between the model and experimental results from Hawkins et al.\cite{hawkins2013high}, and statistical predictions of the model. All profiles are evaluated at 99\% genome replication at $23^\circ \mathrm{C}$.
(a) Total number of active replication forks over time: model versus experiment.
(b) Fraction of genome replicated over time: model versus experiment.
(c) Probability density $P_{io}$ of inter fired origin distances: model versus experiment.
(d) Spatiotemporal replication profile (STRP) of chromosome 14: model versus experiment (Pearson correlation 0.65, $p<0.001$). Shaded region indicates the standard deviation of the ensemble-averaged simulations. 
(e) Distribution of total replication times ($T_{rep}$) from 1000 simulations, revealing a pronounced right-skewed tail. These rare, large time-consuming events events arise only when stochastic variability in fork speed is included, identifying fork-speed heterogeneity as a key determinant of replication-time variability.}
\label{fig:validation}
\end{figure*}

Existing stochastic models in yeast or humans operate at high resolution ($\sim$ 1 kbp) \cite{berners2025regulation, hawkins2013high, berkemeier2025dna, lob20163d}, which means that replication dynamics are averaged over coarse-grained genomic units. As a result, individual fork trajectories and nucleotide-level replication histories are not explicitly resolved, which restricts detailed analysis of fine-scale replication dynamics and localized mutation patterns. In contrast, our lattice-based framework achieves single–base pair resolution (1 bp), explicitly tracking replication timing and fork propagation across the entire genome at the highest physically meaningful resolution. This fine-resolution approach enables precise quantification of replication kinetics and provides a natural platform for investigating local DNA replication under point mutations or clustered point mutation events. All DNA replication results presented here are averaged over 1000 independent ensemble realizations, ensuring robust statistical characterization of replication dynamics.
\section*{\textbf{Results}}
\subsection*{Benchmarking the model}
 Based on the experimental data of the active forks by Hawkins et al. \cite{hawkins2013high}, the maximum value of the limiting factor is set to $L_{max} = 250$ in our simulation. Accordingly, L(t) varies dynamically over time. Theulot et al. \cite{theulot2022genome} demonstrate fork speed distribution at $25^\circ\mathrm{C}$
 and $30^\circ\mathrm{C}$ for the BT1 strain of \textit{S. cerevisiae}. However, we require  parameters characterizing the distribution ($\mu$, $\omega$, $\alpha$ ) at $23^\circ\mathrm{C}$ in order to validate our results against the experimental data of Hawkins et al. \cite{hawkins2013high}. Using a linear extrapolation of these parameters from $25^\circ\mathrm{C}$ and $30^\circ\mathrm{C}$, we obtain a skewness ($\alpha=-1$), width ($\omega=630$), and mean ($\mu=1528$ bp/min) for  
 $23^\circ\mathrm{C}$ (SI). Notably, the extrapolated mean value is very close to the experimentally observed mean fork speed ($\approx$1600 bp/min) and consistent with a total replication time of  $T_{rep} = 83$ minutes at $23^\circ\mathrm{C}$ reported by Hawkins et al. \cite{hawkins2013high}. This motivates us to investigate DNA replication under various thermal stresses using linear extrapolation to characterize parameters of the fork speed distribution, as discussed in the subsection on DNA replication dynamics in \textit{S. cerevisiae} under Thermal Stress.

At the beginning of the S phase, the value of L(t) is low, which restricts origin firing (Fig.~\ref{fig:S1}), consequently resulting in only a small number of active forks F(t) (Fig.~\ref{fig:validation}:a) replicating the genome. This limitation results in a shallow slope of the replicated-fraction curve, leading to a slow increase over time (Fig.~\ref{fig:validation}:b). As the S-phase progresses, L(t) gradually increases, allowing more origins to fire and causing a steady rise in active forks, and the replicated fraction curve. As L(t) approaches saturation, the origin firing reaches its maximum, accompanied by a corresponding peak in the number of active replication forks. This results in a sharp transition in the replicated fraction, characterized by a change from a low to a steeply increasing slope consistent with experimental observations (Fig.~\ref{fig:validation}:b), during which DNA replication proceeds at its fastest rate. As the S phase progresses, an increasing number of potential origins either fired or become passively replicated, leading to a progressive decrease in origin firing intensity (Fig.~\ref{fig:S1}). No new origins fire during the second half of the S phase (Fig.~\ref{fig:S1}), and replication continues only through the already available active forks F(t). Over time, these forks begin to merge with others approaching from the opposite direction, reducing the number of active forks (Fig.~\ref{fig:validation}:a). As fork merging increases, the overall replication rate declines, causing the replicated fraction curve to gradually level off toward saturation (Fig.~\ref{fig:validation}:b). 

We also compute the probability density `$P_{io}$' of inter-fired origins after the completion of S-phase, which can be interpreted as a measure of origin strength: a larger inter-fired origin distance indicates that the corresponding origin replicated more base pairs through its daughter forks (Fig.~\ref{fig:validation}:c). The tail of the  $P_{io}$ distribution extends up to 200 kbp, in close agreement with experimental observations. However, the spatiotemporal replication profile (STRP) provides quantitative genome-wide replication time. For simplification, base pairs are marked at regular intervals of 1000 bp, and the replication time $t_{rep}$ of each marked base pair is recorded. We validate our model using the STRP of chromosome 14 (Chr 14), for which experimental data were reported by Hawkins et al. \cite{hawkins2013high} (Fig.~\ref{fig:validation}:d). The Pearson's correlation coefficient of 0.65, with a p-value much less than 0.0001, indicates a strong correlation between the experiment and simulation STRP (Fig.~\ref{fig:validation}:d). 

In addition, we benchmark the model against multiple experimental datasets from independent studies, demonstrating its generality. The activation function $A_f$, defined as the ratio of the number of fired origins to the number of unreplicated base pairs at a given time(Fig.~\ref{fig:S1}). The correlation between active fork density and origin firing intensity is compared with Ma et al.\cite{ma2012replication}(Fig.~\ref{fig:S1}).

Though there is a degree of deterministic ordering in the spatial distribution of potential origins, origin firing is dominated by temporally stochastic activation. In addition to stochastic origin firing, the model incorporates variability in fork speeds, drawn from a skewed Gaussian distribution rather than assuming a single mean value. Together, origin firing and fork progression complete DNA replication within a total replication time, `$T_{rep}$'. The histogram of
`$T_{rep}$' obtained from 1000 simulations is shown in (Fig. \ref{fig:validation}:e), with a mean of 83 min. The distribution is consistent with an Erlang form, exhibiting a tail extending up to 200 min. This behavior arises from replication dynamics driven by a subset of slow-moving forks that fail to merge efficiently. These forks, with speeds significantly below the mean, contribute to rare `$T_{rep}$' values that are substantially longer than average. For comparison, simulations performed with a constant fork speed of 1526 bp/min (equal to the mean of the distribution) yield a $T_{rep}$ distribution (Fig. \ref{fig:S1}) that is approximately Gaussian with a narrow variance, consistent with previous reports by Berners-Lee et al.\cite{berners2025regulation}, and unable to account for long S-phase durations. These results demonstrate that heterogeneity in fork speed is essential for generating rare but biologically significant long-replication events. Such an Erlang distribution has been reported for doubling-time distributions in E. coli\cite{pugatch2015greedy} and for S-phase duration in RPE, U2OS, and H9 human cell lines\cite{chao2019evidence}. However, to our knowledge, neither experimental studies nor theoretical models incorporating fork-speed distributions have reported the S-phase duration distribution in \textit{S. cerevisiae}. By incorporating a more realistic representation of fork-speed variability, our model demonstrates that the S-phase duration in \textit{S. cerevisiae} also follows an Erlang distribution.
\subsection*{DNA replication dynamics in \textit{S. cerevisiae} under Thermal Stress}
\begin{figure*}[bth]
\centering
\includegraphics[scale=0.65]{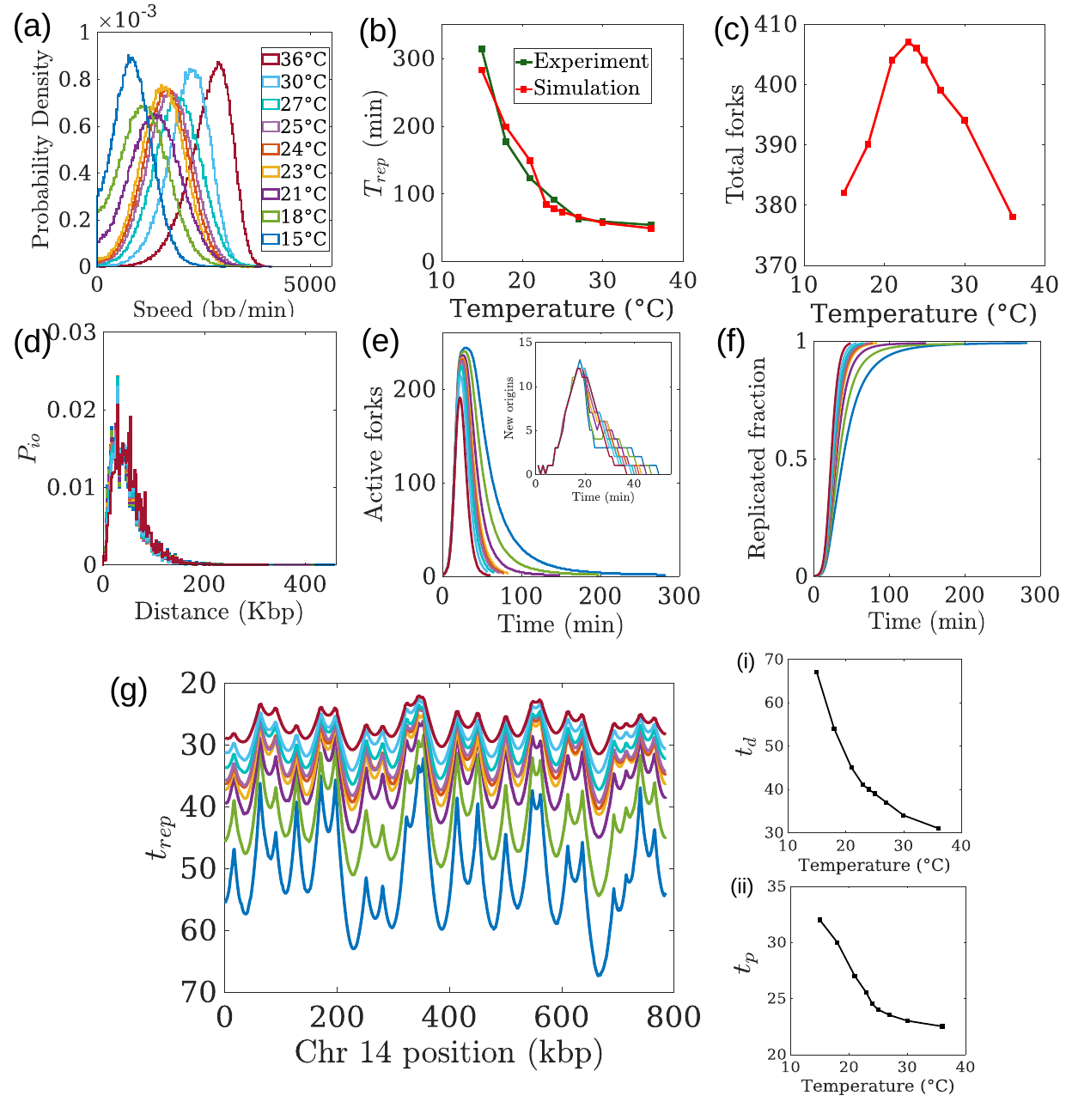}
\caption{DNA replication dynamics under thermal stress.
(a) Probability density distribution of replication fork speeds displays a skewed Gaussian profile. Linear extrapolation of experimentally measured distributions at ($25^\circ\mathrm{C}$) and ($30^\circ\mathrm{C}$) (SI) determines the temperature dependence of the skewness ($\alpha$), width ($\omega$), and mean fork speed ($\mu$) (SI).
(b) The extrapolated fork speed distributions accurately reproduce the experimentally measured total replication time ($T_{\mathrm{rep}}$) across all temperatures, supporting the robustness of the thermal extrapolation framework.
(c) Total number of replication forks generated during S-phase shows a subtle yet significant non-monotonic dependence on temperature. Even small changes in the number of forks remain biologically important, as each additional stalled or collapsed fork increases the risk of DNA damage and cell death under thermal stress.
(d) Probability density ($P_{io}$) of inter-fired origin distances under different thermal stress conditions. The distribution develops a stronger long tail at lower temperatures, with the highest tail at ($15^\circ\mathrm{C}$), followed by ($18^\circ\mathrm{C}$), indicating enhanced origin strength.
(e) Temporal evolution of active fork number ($F(t)$). The inset shows the rate of new origin firing ($I(t)$), where the peak firing rate ($I_0$) remains nearly invariant across temperatures, while additional late-origin firing events appear at lower temperatures.
(f) Total replicated genome fraction as a function of time. During late S-phase, the replication rate (slope) decreases markedly at lower temperatures, resulting in a substantial increase in ($T_{\mathrm{rep}}$). 
(g) STRP of chromosome 14 under different thermal stress conditions. Replication timing ($t_{\mathrm{rep}}$) preserves its global spatial organization across temperatures, while both the minimum replication timing $(t_p)$ of early-replicating regions and the maximum replication timing $(t_d)$ of late-replicating regions increase significantly with decreasing temperature (i-ii).}
\label{fig:temp}
\end{figure*}

Temperature is a key environmental determinant of \textit{S. cerevisiae} proliferation, with optimal growth occurring between 25–32 °C and peak rates observed near 30°C ~\cite{vanoni1984effects}. As cell proliferation is tightly coupled to DNA replication dynamics, quantifying replication fork progression under these conditions is essential. However, wild-type
\textit{S. cerevisiae} can not incorporate the nucleotide analogue BrdU, which is commonly used to study DNA replication dynamics. Theulot et al. \cite{theulot2022genome} developed the BT1 strain, which retains wild-type growth characteristics while efficiently incorporating BrdU, and reported a distribution of replication fork speeds at $25^\circ\mathrm{C}$ and $30^\circ\mathrm{C}$.  The characteristic parameters of the fork-speed distribution ($\alpha$, $\omega$, and $\mu$) at $25^\circ\mathrm{C}$ and $30^\circ\mathrm{C}$ are extracted from Theulot et al. \cite{theulot2022genome} as discussed in the `Formulation of Fork speed distribution' subsection. Based on the characteristic parameters derived from fork speed distributions at $25^\circ\mathrm{C}$ and $30^\circ\mathrm{C}$, we perform a linear extrapolation to estimate parameter values at additional temperatures such as $15^\circ\mathrm{C}$, $18^\circ\mathrm{C}$, $21^\circ\mathrm{C}$, $23^\circ\mathrm{C}$, $24^\circ\mathrm{C}$, $27^\circ\mathrm{C}$, and $36^\circ\mathrm{C}$ (Fig.~\ref{fig:S2}) and construct the corresponding fork-speed distribution (Fig.~\ref{fig:temp}:a). The temperature specific fork-speed distributions (Fig.~\ref{fig:temp}:a), together with stochastic origin firing, accurately recapitulate the total replication time, $T_{rep}$, across varying thermal stress conditions (Fig.~\ref{fig:temp}b) without adjustment of any additional model parameters. The predicted $T_{rep}$ values closely match experimentally observed S-phase durations reported by Vanoni et al.~\cite{vanoni1984effects}, indicating that thermal stress primarily modulates replication dynamics through fork velocity while leaving the availability of the limiting replication factor largely unchanged.

We further find that the total number of forks generated during the S phase exhibits a non-monotonic dependence on temperature (Fig.~\ref{fig:temp}:c). The total fork number increases with temperature up to approximately $23^\circ\mathrm{C}$, where it reaches a maximum, and then decreases as the temperature is further increased. This behavior can be understood from the temperature dependence of fork speed and origin activity. The probability density of inter-fired origins, $P_{io}$, (Fig.~\ref{fig:temp}:d) reflects the effective strength of replication origins. The tail of the  $P_{io}$ distribution is longest at $15^\circ\mathrm{C}$, followed by $18^\circ\mathrm{C}$, indicating stronger origin strength to replicate at lower temperatures. The inset of (Fig.~\ref{fig:temp}:e) shows the temporal evolution of the origin firing I(t). Although the maximum firing rate $I_{max}$ remains nearly identical across all thermal stress conditions, the maximum number of active forks $F_{max}$ is consistently higher at lower temperatures than at higher temperatures, showing a monotonic order (Fig.~\ref{fig:temp}:e).

Hence the slower fork speed, higher strength of origins and the high value of $F_{max}$ at lower temperature, but almost the same  $I_{max}$ value for all temperature conclude that at lower temperatures like $15^\circ\mathrm{C}$, slower fork progression leads to delayed fork merging, resulting in a larger number of simultaneously active forks (Fig.~\ref{fig:temp}:d). These persistent forks sequester limiting factors, thereby suppressing additional origin firing and reducing the total number of forks generated. As the temperature increases from $15^\circ\mathrm{C}$ to $23^\circ\mathrm{C}$, fork speed increases approximately linearly, leading to more frequent fork merging and the release of limiting factors, which in turn promotes additional origin firing and increases the total number of forks. Beyond $23^\circ\mathrm{C}$, fork speeds become sufficiently high that many origins are passively replicated before firing, resulting in fewer fired origins, and a reduced total number of forks, as shown in (Fig.~\ref{fig:temp}:c).

The progression of genome replication is illustrated by the replicated fraction profiles in (Fig.~\ref{fig:temp}:f). At higher temperatures $\sim 36^\circ\mathrm{C}$, the slope of the replicated-fraction curve begins to decrease only when replication is nearly complete (~98\%). In contrast, at lower temperatures, the slope decreases at earlier times, resulting in an overall reduction in the DNA replication rate and longer S phase duration driven by slower fork speed. The spatiotemporal replication profile (STRP) under thermal stress for chromosome 14 is shown in (Fig.~\ref{fig:temp}:g).  (Fig.~\ref{fig:temp}:g:I) shows the maximum replication time, denoted as $t_{d}$, which corresponds to the most late-replicating region of chromosome 14 as a function of temperature. Similarly, (Fig.~\ref{fig:temp}:g:II) shows the earliest replication time, denoted as $t_{p}$, corresponding to the earliest replicating region of chromosome 14. Both quantities decrease monotonically with increasing temperature, indicating that all regions of chromosome 14 replicate more rapidly at higher temperatures. At lower temperatures, up to approximately $23^\circ\mathrm{C}$, the late replication time $t_{d}$ is nearly twice the early replication time $t_{p}$. In contrast, at higher temperatures (above ~ $25^\circ\mathrm{C}$), $t_{d}$ is approximately 1.5 times $t_{p}$. In addition, as the temperature decreases, the temporal separation between neighboring peaks and troughs in the STRP increases, leading to sharper and more widely spaced replication-timing features. However, the STRP pattern remains unchanged across temperatures; as a whole, it appears to shift by a scaling factor and stretch between peaks and troughs.

\subsection*{DNA replication dynamics in \textit{S. cerevisiae} under Chemical Stress}

Hydroxyurea (HU) is one of the most commonly used drugs to study DNA replication under chemical stress. It induces replication stress in \textit{S. cerevisiae} by inhibiting the enzyme ribonucleotide reductase (RNR), an enzyme that catalyses the conversion of ribonucleotides to deoxyribonucleotides (dNTPs) for DNA replication and repair, and hence depleting the dNTP pool and decreasing the fork progression rate \cite{shaw2024revised}. Efficient DNA replication depends on a constant supply of dNTPs, which are produced throughout the S-phase. There is a tight control over the production of dNTPs, as deregulated or imbalanced dNTP pools are deleterious for genome integrity \cite{kumar2010highly}
This control is mostly exerted by RNR and it serves as a rate-limiting step in dNTP biosynthesis. \\

\begin{figure*}[t]
\centering
\includegraphics[scale=0.7]{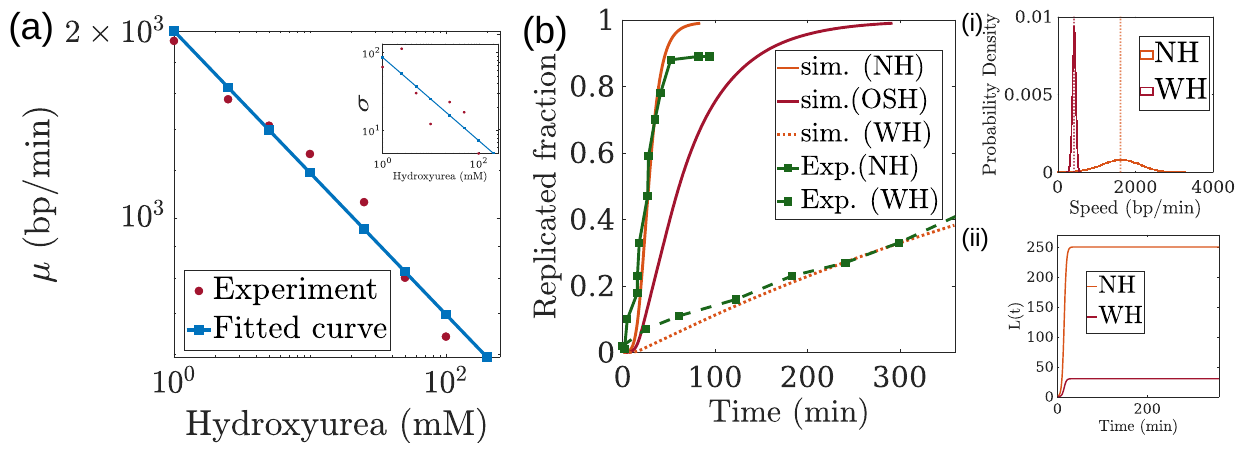}
\caption{DNA replication under chemical stress (Hydroxyurea). 
(a) Log–log linear fit of the mean replication fork speeds ($\mu$) as a function of Hydroxyurea (HU) concentrations reveals a robust power-law dependence, which enables estimation of ($\mu$) at 200 mM HU. The inset shows the corresponding standard deviation ($\sigma$). All results are obtained at ($30^\circ\mathrm{C}$).
(b) Replication dynamics at ($23^\circ\mathrm{C}$) under 200 mM HU. Comparison between NH (No HU), WH (With HU), and OSH (Only fork speed affected by HU) demonstrates that HU-induced replication stress cannot be explained solely by reduced fork speed (i), highlighting the critical contribution of origin regulation through limiting factor (ii) and global replication-program remodeling under chemical stress.
}
\label{fig:drug}
\end{figure*}

Experiments by Theulot \textit{et al} \cite{theulot2022genome} on the BT1 strain measured the mean replication fork speed `$\mu$', and standard deviation `$\sigma$' at $30^\circ\mathrm{C}$ for various HU concentrations (1, 2.5, 5, 10, 25, 50, and 100 mM). We obtain those experimental values from Theulot et al. \cite{theulot2022genome} and analyze them on a log–log scale (Fig.~\ref{fig:drug}:a). The data exhibit an approximately linear trend, consistent with a power-law correlation as $\mu=A_{\mu}[HU]^{\beta_{\mu}}$ and  $\sigma=A_{\sigma}[HU]^{\beta_{\sigma}}$. Here, [HU] denotes the hydroxyurea concentration in mM and coefficients, $A_{\mu}$ = 2024, $A_{\sigma}$ = 86.75, with corresponding exponents  $\beta_{\mu}$ = -0.23 and $\beta_{\sigma}$ = -0.54.

Although in fork speed distribution the skewness parameter $\alpha$ varies under thermal stress~\cite{theulot2022genome}, there are currently no experimental evidence indicating a change in $\alpha$ under HU treatment. Since all HU experiments are performed at a fixed temperature of $30^\circ\mathrm{C}$, we therefore keep `$\alpha=-2$' constant. Under this assumption, a linear proportionality between the standard deviation $\sigma$ and the width parameter $\omega$ follows naturally. The corresponding values of $\omega$ are then calculated using the proportionality constant, yielding the fork-speed distributions for all HU concentrations (Fig.~\ref{fig:S3}).

Using the power-law correlation and the corresponding proportionality constants, we extrapolate $\mu$, $\sigma$, (Fig.~\ref{fig:drug}:a) and $\omega$ respectively to an HU concentration of 200 mM at $30^\circ\mathrm{C}$. At this concentration, HU reduces the mean fork speed $\mu$ by approximately a factor of 3.6, from 2128 bp/min to 590 bp/min. We consider HU stress to act independently of thermal stress. Accordingly, at $23^\circ\mathrm{C}$, the mean fork speed under 200 mM HU (after applying the same 3.6-fold reduction) is estimated to be approximately 425 bp/min. However, experimental measurement of the standard deviation $\sigma$ at 200 mM HU and $23^\circ\mathrm{C}$ is not available. In the absence of additional experimental constraints, we therefore use the same value of the parameter $\sigma$ as measured at 200 mM HU and $30^\circ\mathrm{C}$, accordingly measure $\omega$. Because $\omega$ is already small at $30^\circ\mathrm{C}$ and is expected to decrease further at lower temperatures due to the explained power law correlation, this approximation is unlikely to affect our conclusions. Using a temperature-dependent skewness parameter, $ \alpha=-1$, the corresponding $\omega$, and a mean fork speed of 425 bp/min at 200 mM HU and $23^\circ\mathrm{C}$, we obtain the fork-speed distribution (Fig.~\ref{fig:drug}:b:I).

Experiments by Theulot et al.~\cite{theulot2022genome} characterize the effect of HU on the fork-speed distribution, but do not provide information on changes in the availability of limiting factors. When only the HU-induced fork speed is considered, the resulting replicated-fraction profile `OSH'(Fig.~\ref{fig:drug}:b), does not agree with the experimentally observed profile reported by Alvino et al.~\cite{alvino2007replication}. This discrepancy indicates that additional factors must contribute to the replication stress response under HU treatment. In particular, the availability of limiting factors that regulate origin firing plays a crucial role. Nonetheless, a qualitative reduction of origin firing has been reported by Alvino et al.~\cite{alvino2007replication} under HU treatment. By reducing limiting factor to 30\% (Fig.~\ref{fig:drug}:b:II) of its original value `250' ~\cite{hawkins2013high}, the replicated-fraction profile produced by our model closely matches the experimental results of Alvino et al.~\cite{alvino2007replication} at $23^\circ\mathrm{C}$ under 200 mM HU (Fig.~\ref{fig:drug}:b). These results suggest that HU affects not only the fork speed distribution but also the limiting factor. Furthermore, the model provides a quantitative prediction of the HU-dependent modulation of the limiting factor.

Since chemical and thermal stresses are assumed to act independently, we use L(t) obtained at $30^\circ\mathrm{C}$ for 0 mM HU and L(t) obtained at  $23^\circ\mathrm{C}$ for 200 mM HU as reference cases. A linear interpolation between these two limits yields the limiting-factor profiles for intermediate HU concentrations Fig.~\ref{fig:S3}.
Using the fork-speed distributions (Fig.~\ref{fig:S3}) together with the corresponding limiting factor profiles we compute the time evolution of the replicated fraction for each HU concentration (specifically those for which experimental values of $\mu$ and $\sigma$ are reported by Theulot et al.~\cite{theulot2022genome}(Fig.~\ref{fig:S3}). We also explore the spatiotemporal replication profile (STRP) under HU stress for chromosome 14 (Fig.~\ref{fig:S3}).
Despite the substantial HU-induced changes in fork speed and in the limiting factor function L(t), the overall pattern of the STRP does not change significantly. Instead, the profile appears to be largely preserved and uniformly shifted in time, consistent with experimental observations reported by Alvino et al.~\cite{alvino2007replication}.

\subsection*{DNA replication dynamics in \textit{S. cerevisiae} under Genotoxic Stress}

In this section, we consider DNA replication in  \textit{S. cerevisiae} under conditions of genotoxic stress, as reported by Theulot et al. in the BT1 strain. This strain exhibits growth properties similar to the wild-type strain and efficiently incorporates BrdU into the genome, enabling measurement of the mean replication fork speed and its standard deviation for five different BT1 mutant strains \cite{theulot2022genome}.
\begin{figure*}[t]
\centering
\includegraphics[scale=0.6]{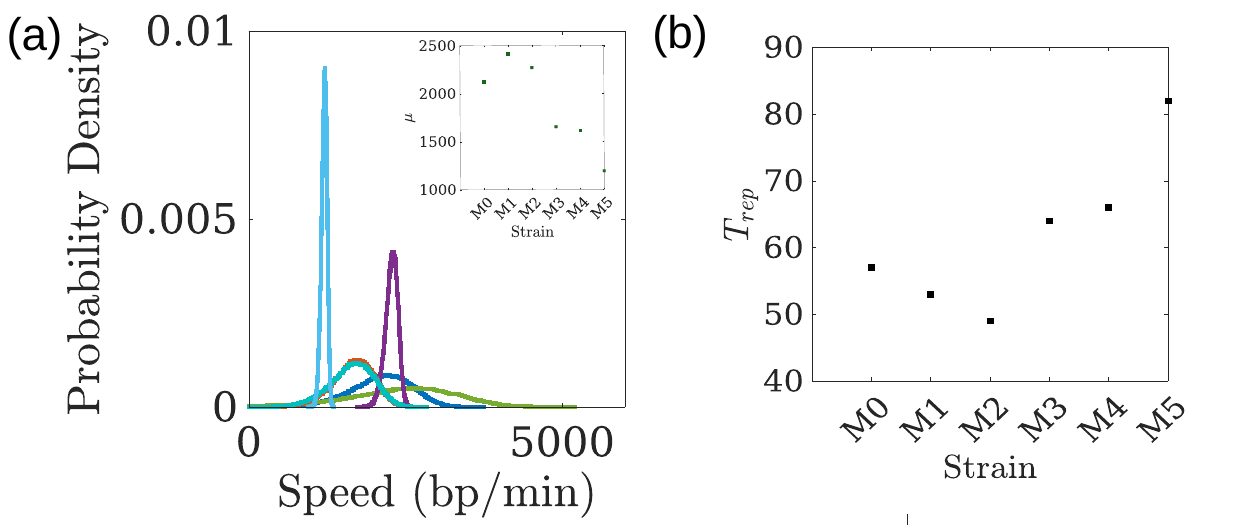}
\caption{DNA replication under genotoxic stress. M0: BT1 (wild type); M1: BT1 rtt109$\Delta$; M2: BT1 sml1$\Delta$; M3: BT1 csm3$\Delta$; M4: BT1 tof1$\Delta$; M5: BT1 mrc1$\Delta$.
(a) Fork-speed distributions at $30^\circ\mathrm{C}$ for all strains, with insets showing experimentally observed mean fork speeds.
(b) Total replication time $(T_{\mathrm{rep}})$ for different genotoxic strains.}

\label{fig:genotoxic}
\end{figure*}
Since these experiments were performed at $30^\circ\mathrm{C}$, we keep the skewness parameter $\alpha=-2$ fixed for all strains. The width parameter $\omega$ is calculated for each strain using the linear relationship with $\sigma$ as  $\alpha$ is constant. Using the parameters $\mu,\omega,\alpha$, we obtain the fork-speed distributions for all strains (Fig.~\ref{fig:genotoxic}). 

The first mutant, M1 (BT1 rtt109$\Delta$) lacks the histone acetyltransferase, Rtt109 that acetylates histone H3 before its binding to the DNA. Histone modifications, such as acetylation, regulate nucleosome stability by promoting proper assembly of the nucleosome as the fork progresses. This imposes a controlled constraint on the fork movement. In the absence of Rtt109, defective chromatin reassembly reduces this constraint, allowing forks to move faster but at the cost of increased replication stress and genome instability \cite{frenkel2021rtt109}. In addition to increase in fork speed, the M1 strain has been reported to suppress the limiting factor function L(t) by approximately 15\%, compared to the wildtype strain (M0) \cite{frenkel2021rtt109}, whereas the other strains do not exhibit a significant effect on L(t). Despite the slight increase in fork speed, literature suggests no significant changes in this mutant with respect to the duration of S-phase, majorly owing to the increase in replicon lengths to compensate for the speed changes \cite{frenkel2021rtt109, kaplan2008cell, ide2013rtt109, han2007acetylation}. 

M2 strain (BT1 sml1$\Delta$) lacks Sml1, an allosteric inhibitor of ribonucleotide reductase. It acts during the G1 phase of the cell cycle to limit dNTP sysnthesis and prevent premature DNA replication \cite{zhao2001ribonucleotide}. Its deletion elevates intracellular dNTP pools, thereby facilitating more rampant DNA synthesis at replication forks and increasing the fork speed \cite{poli2012dntp} \cite{mantiero2011limiting} The fork speed for this mutant shows a much narrower distribution with similar mean fork speed as compared to M0 (Fig. 6a). This might be indicative of the possibility that forks that were earlier progressing slowly start replicating much faster owing to the higher availability of dNTPs. This also allows slightly early completion of DNA replication, as shown in (Fig.~\ref{fig:genotoxic}).

Csm3, a component of the fork protection complex is deleted in M3 strain (BT1 csm3$\Delta$). Csm3 is a replisome-associated factor that stabilises replication forks and regulates checkpoint signalling during DNA replication. Deletion of Csm3, a component of the Csm3–Tof1 fork protection complex, compromises stable association of Mrc1 with the replisome and consequently reduces replication fork progression efficiency \cite{yeeles2017eukaryotic}. M4 strain (BT1 tof1$\Delta$) lacks Tof1, another component of the fork protection complex which is also a replisome-associated factor that partners with Csm3 to stabilise replication forks and regulate checkpoint signalling during DNA replication. Deletion of Tof1 disrupts the fork protection complex and weakens replisome integrity, leading to reduced replication fork progression efficiency. Consequently, replication forks slow down and become prone to pausing and instability \cite{yeeles2017eukaryotic}. Both of these strains show a shift in the fork speed distribution in Fig. 6a with markedly reduced mean fork speed values. This also results in a moderate increase in the total replication timing as shown in Fig. 6b. The STRPs obtained for M3 and M4 seem to be similar to M0, as shown in (Fig.~\ref{fig:genotoxic}).

M5 strain (BT1 mrc1$\Delta$) lacks Mrc1, a protein crucial for fork protection and stability. In a normal S phase, Mrc1, along with Tof1, loads onto DNA soon after initiation and traverses with replication forks, placing them in a unique position to detect replication stress. Deletion of Mrc1 compromises fork stabilisation and checkpoint signalling, causing cells to progress slowly through S phase and accumulate spontaneous DNA damage \cite{tourriere2005mrc1}. The fork speed distribution for this mutant is drastically affected, with a sharp decline in the mean fork speed as shown in (Fig.~\ref{fig:genotoxic}).  Similarly, the total replication time for M5 is the highest owing to slower fork speeds (Fig.~\ref{fig:genotoxic}). 

 \subsection*{Conclusion}
Moving beyond the conventional picture of deterministic origin firing and
constant fork speed in \textit{S. cerevisiae}, our model incorporates
stochastic origin firing, a key feature of eukaryotic replication dynamics, and
a distributed fork speed landscape. In addition, we introduce a time-dependent
limiting factor that represents the cellular resources controlling origin firing.
Together, these elements constitute a minimal, physics-based
framework, characterized by only two effective degrees of freedom, that
quantitatively captures eukaryotic DNA replication across thermal, chemical, and
genotoxic stress conditions without invoking condition-specific regulatory
mechanisms.

Our results provide a physical interpretation of how global replication dynamics
emerge from competition for limiting replication resources. In particular,
saturation of the limiting factor naturally produces a sharp transition to rapid
genome replication, offering a mechanistic explanation for the experimentally
observed acceleration of S-phase progression. In addition, we demonstrate that
incorporating a distribution of replication fork speeds generates a broad spectrum
of total S-phase durations that follows an Erlang distribution, indicating that
heterogeneity in fork speed is essential for producing rare yet biologically
significant long-replication events. Beyond reproducing known experimental observations, the
framework yields several predictions: a non-monotonic dependence of total fork
number on temperature, a power-law relationship between fork speed and
hydroxyurea concentration, and the requirement to modulate the limiting factor---
not only fork speed---to account for HU-induced replication stress.

The model considers a limited set of origins for which the location, competency,
mean firing time, and width of the probability curve are available. In contrast,
approximately 829 potential origins have been reported in OriDB, and more recent
studies have suggested that the number of potential origins may range from
$\sim$1,600 to $\sim$5,000 \cite{foss2024identification}. The effect of this
limitation is visible in the activation function discrepancy noted in the SI, and
incorporating a more complete origin database is a natural next step. Notably,
the model operates at single base-pair resolution, opening the
framework to investigate replication dynamics in local genomic regions harboring
point mutations or clustered point mutations associated with chronic diseases,
such as Sickle cell disease, Beta thalassemia, and cancers driven by mutation
hotspots, including Colorectal cancer and Breast cancer. Overall, this work
establishes a quantitative foundation for understanding how genome replication
adapts to stress and provides a versatile platform for integrating future
experimental data with theoretical modeling.

\begin{acknowledgments}
A.G. acknowledges financial support from SERB-DST (India) through Projects No.~MTR/2022/000232 and No.~CRG/2023/007056-G. A.G. also acknowledges support from DST (India) through Grants No.~DST/NSM/R\&D\allowbreak\_HPC\allowbreak\_Applications/2021/05 and No.~SR/FST/PSI-215/2016, as well as the Indian Institute of Technology Hyderabad through Seed Grant No.~IITH/2020/09.
\end{acknowledgments}

\bibliography{pnas-sample}

\clearpage
\onecolumngrid


\renewcommand{\thefigure}{S\arabic{figure}}
\renewcommand{\thetable}{S\arabic{table}}
\renewcommand{\theequation}{S\arabic{equation}}
\renewcommand{\thepage}{S\arabic{page}}

\setcounter{figure}{0}
\setcounter{table}{0}
\setcounter{equation}{0}
\setcounter{page}{1}

\section*{Supplementary Information}

\begin{figure}[h]
    \centering
    \includegraphics[width=0.9\textwidth]{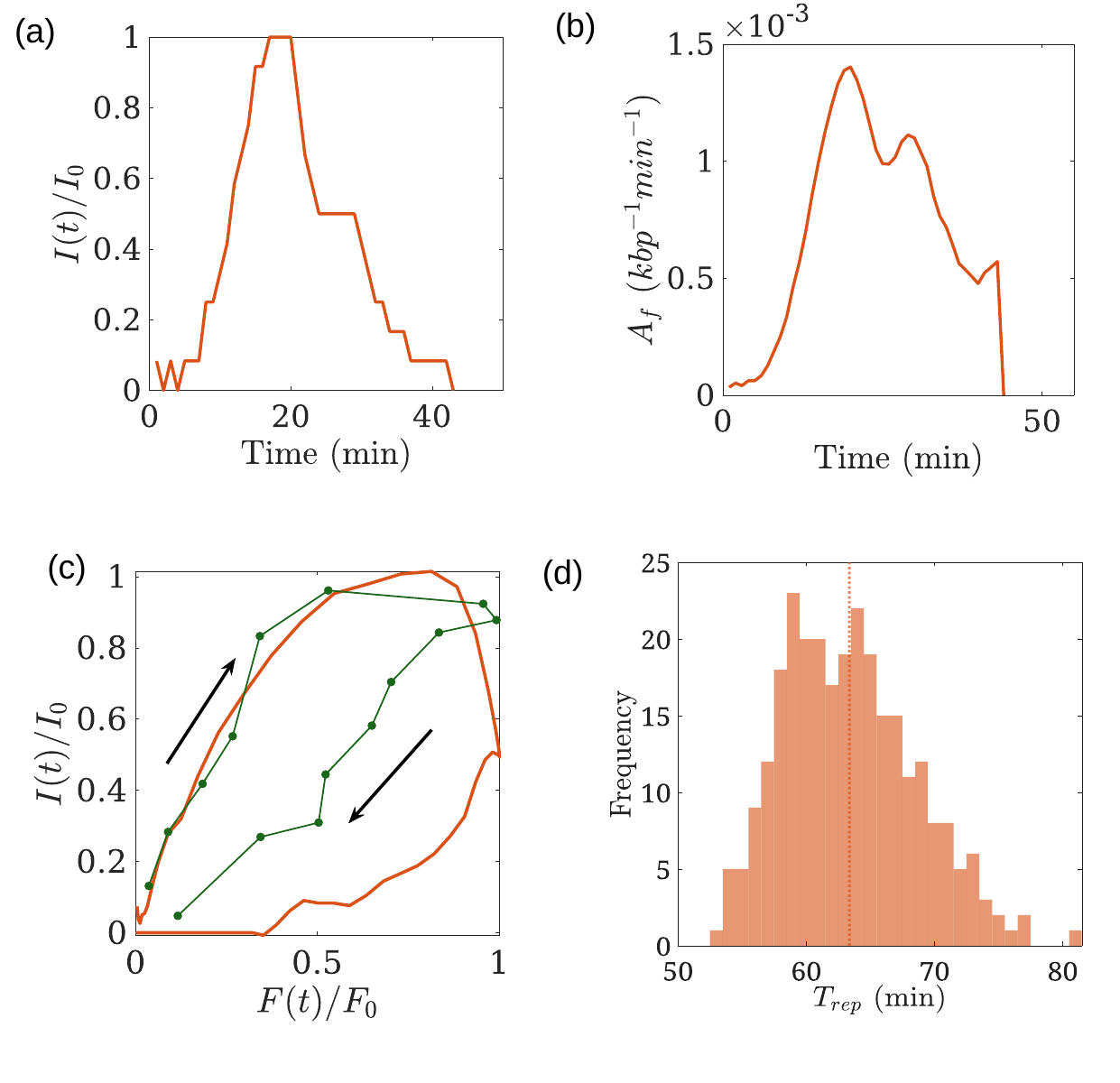}
    \caption{
    (a) Time evolution of fired-origin intensity.
    (b) Activation function $A_f$, defined as the ratio of the number of origins fired at a given time to the number of unreplicated base pairs at that time.
    (c) Correlation between normalized firing-origin intensity $I(t)/I_{0}$ and normalized fork intensity $F(t)/F_{0}$ throughout S phase.
    (d) Frequency distribution of total replication times ($T_{\mathrm{rep}}$) obtained from 1000 simulations assuming a constant fork speed of 1528 bp/min.
    }
    \label{fig:S1}
\end{figure}

\begin{figure}[h]
    \centering
    \includegraphics[width=0.9\textwidth]{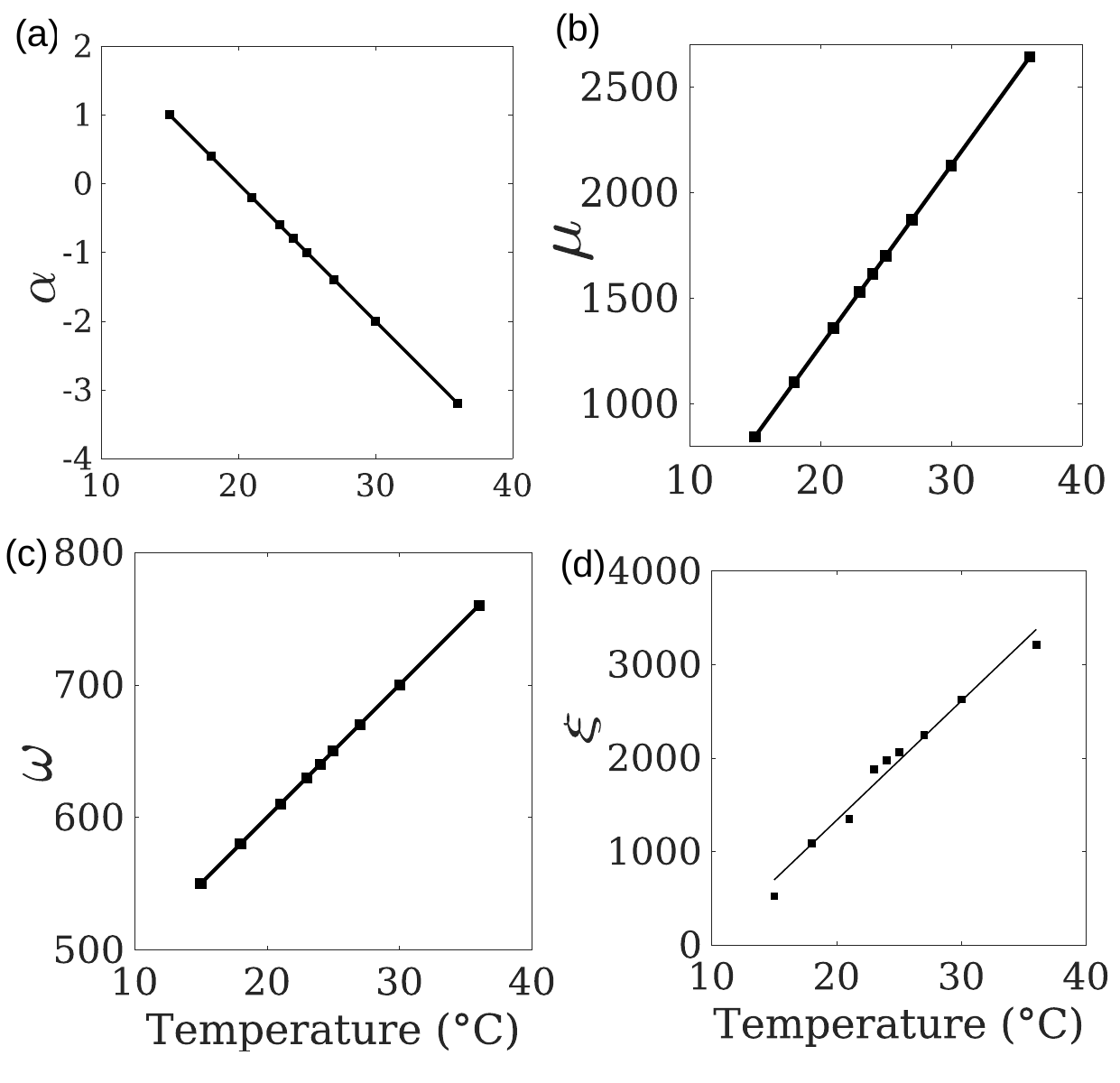}
    \caption{
    (a-c) Linear extrapolation used to estimate skewness ($\alpha$), width ($\omega$), and mean ($\mu$) at different temperatures based on experimental data at $25^\circ$C and $30^\circ$C.
    (d) The resulting extrapolated location parameter ($\xi$) varies linearly with temperature.
    }
    \label{fig:S2}
\end{figure}

\begin{figure}[h]
    \centering
    \includegraphics[width=0.9\textwidth]{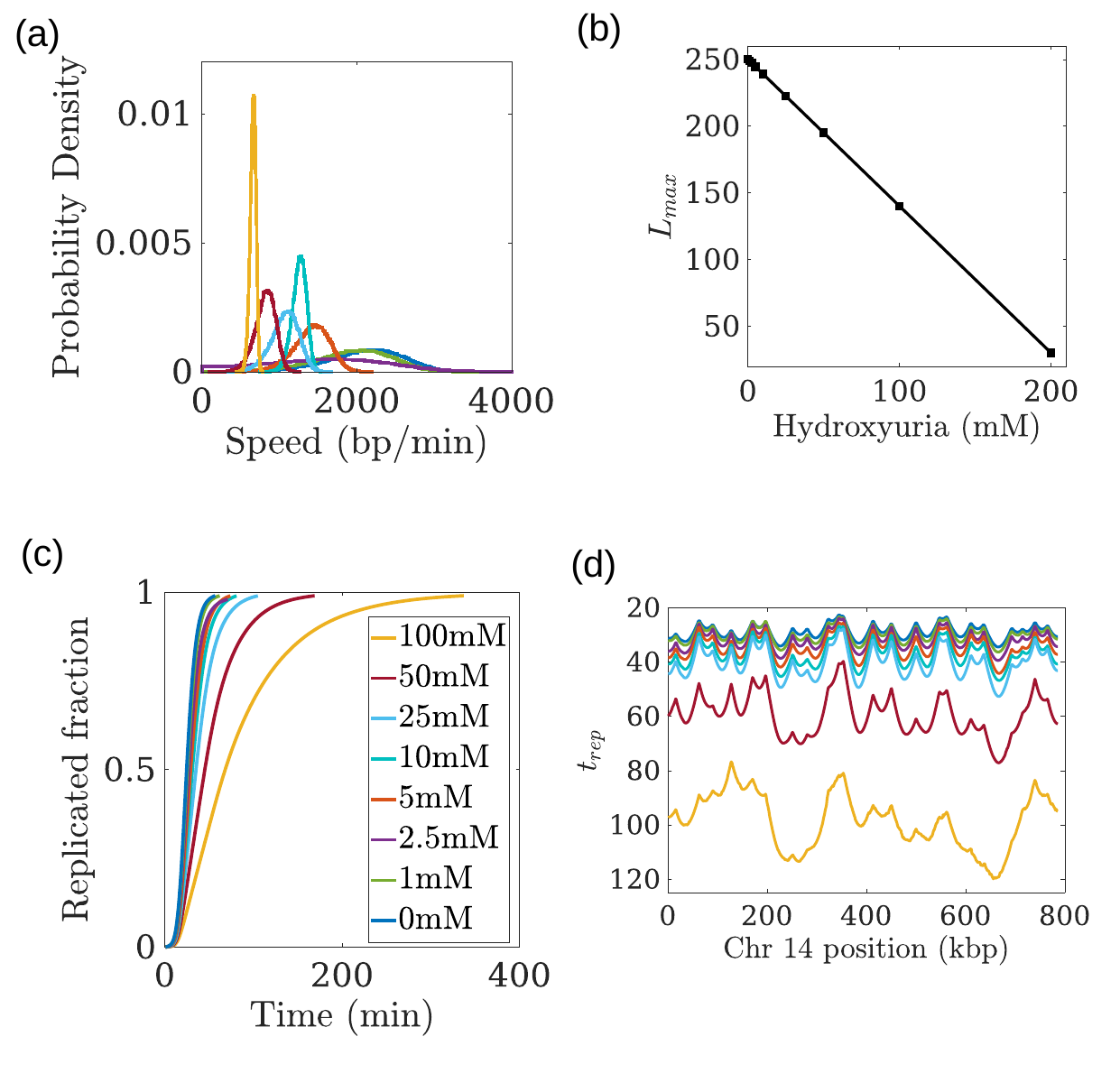}
    \caption{
    (a) Fork-speed distributions for different HU concentrations.
    (b) Assuming the effect of chemical stress (HU) is independent of temperature, $L_{\max}$ at intermediate HU concentrations was estimated by linear interpolation between the 0 mM and 200 mM cases.
    (c) Replicated fraction growth with time for 0-100mM HU concentration.
    (d) Replication timing ($t_{\mathrm{rep}}$) largely preserves its spatial pattern under low HU, while higher HU reduces fine-scale structure.
    }
    \label{fig:S3}
\end{figure}

\end{document}